\documentclass[aps,prb,onecolumn,amsmath,notitlepage,floatfix,footinbib,superscriptaddress,showpacs, showkeys]{revtex4-1}
\usepackage[colorlinks=true]{hyperref}%url}
\usepackage{amsmath}
\usepackage{epstopdf}
\usepackage{color}
\usepackage{graphicx}
\usepackage{amssymb}
\usepackage[figuresright]{rotating}
\begin{document}

%\begin{frontmatter}



%\dochead{28th Annual CSP Workshop on ``Recent Developments in Computer Simulation Studies in Condensed Matter Physics'', CSP 2015}
%% e.g. \dochead{17th International Conference on Dynamical Processes in Excited States of Solids}

\title{\texttt{opendf} - an implementation of the dual fermion method for strongly correlated systems }


\author{Andrey E. Antipov} 
\email{aantipov@umich.edu}
\author{James P.F. LeBlanc}
\author{Emanuel Gull}

\affiliation{Department of Physics, University of Michigan, Ann Arbor, Michigan 48109, USA}

\begin{abstract}
The dual fermion method is a multiscale approach for solving lattice problems of interacting strongly correlated systems. In this paper, we present the \texttt{opendf} code, an open-source implementation of the dual fermion method applicable to fermionic single-orbital lattice models in dimensions $D=1,2,3$ and $4$. The method is built on a dynamical mean field starting point, which neglects all local correlations, and perturbatively adds spatial correlations. Our code is distributed as an open-source package under the GNU public license version 2.
\end{abstract}

\maketitle

\section{Introduction}

Understanding the physics of complex correlated electron systems beyond simple approximations or exactly solvable limits is a long-standing goal of condensed matter physics. Towards this goal, the dynamical mean field theory (DMFT) \cite{MetznerVollhardt:1989,MH89,Georges92,Jarrell92,Georges1996,KotliarSavrasov:2006} is a workhorse which provides numerically simulated results for the physics of such systems. It establishes that, if correlations and interactions are assumed to be local, the (intractable) extended system can be mapped self-consistently onto an  Anderson impurity model, which can then be solved numerically.

The dynamical mean field approximation of locality is often precise enough that general material trends can be reproduced. Nevertheless, cases where non-local correlations lead to behavior not captured by DMFT are known \cite{Lichtenstein2000,MaierJarrell:2005,Held2008,Fuchs2011}, and therefore methods that improve on this approximation are needed. The dual fermion method \cite{Rubtsov2008}, which perturbatively adds corrections to a DMFT starting point reintroduces momentum dependent correlations. If all corrections are included, the method recovers the full momentum dependence of the original problem and becomes numerically exact.

In this paper, we present \texttt{opendf}, an implementation of the `ladder series' variant of the Dual Fermion method \cite{HafermannLi:2009}. This variant is approximate, as neither vertices with more than four legs nor series of vertices beyond a single ladder are considered. Nevertheless, it has been shown to consistently improve on DMFT results \cite{Rubtsov2008,Brener2008,HafermannLi:2009,AntipovRubtsov:2011,Li2014,Otsuki2014} and capture critical properties of phase transitions \cite{Antipov2014,Hirschmeier2015}.

Dual fermion calculations rely on a dynamical mean field input, which can be provided by one of the publicly available open source software packages that implement the approximation, including ALPS (the Algorithms and Libraries for Physics Simulations) \cite{ALPS2}, TRIQS (the Toolbox for Research on Interacting Quantum Systems) \cite{TRIQS}, and iQIST \cite{iQIST}.
This initial step requires the self-consistent solution of an interacting quantum many-body system and the calculation of vertex functions \cite{Hafermann2013a,Gull2008b} and is computationally much more expensive than the summation of the dual fermion diagrams. 
%Calculations by \texttt{opendf} therefore can augment any DMFT computation. 

The rest of this paper is organized as follows: Section~\ref{sec:meth} introduces the methodology. Section~\ref{sec:imp} describes  distribution aspects, section~\ref{sec:perf} performance aspects, section~\ref{sec:ex} shows some examples, and section \ref{sec:conclusions} will conclude.

\section{Methodology}\label{sec:meth}
\subsection{Prerequisites}
We consider a general fermionic single-orbital lattice model with a Hamiltonian
\begin{equation}
H = \sum_{k\sigma} (\varepsilon_k - \mu) c^\dagger_{k\sigma} c_{k\sigma} + \sum_i H^{\mathrm{int}} [c^\dagger_i, c_i],
\end{equation}
written in mixed momentum, $k$, and real space, $i$, notation in terms of creation and annihilation operators ($c^\dagger_{k\sigma}$ and $c_{k\sigma}$ respectively).  The index $\sigma$ labels the spin projection, $\varepsilon_k$ is the lattice dispersion relation and $k$ is the vector in the reciprocal space. 
$H^{\mathrm{int}}$ is the local interaction for each site, $i$, on the lattice. 
No assumption is made within DF as to the structure of $H^{\mathrm{int}}$.  %For example, $H^{\mathrm{int}} = U c^\dagger_\uparrow c_\uparrow c^\dagger_\downarrow c_\downarrow$ would refer to the typical Hubbard model \cite{Hubbard1963}. 

As a first step, which must be performed outside of this code, an approximate solution of the model is obtained from a dynamical mean field calculation, for example provided by the ALPS code \cite{ALPS2} with an appropriate impurity solver \cite{Hafermann2013a}.
It provides an estimate for the local Green's function of the lattice problem as a  solution of the Anderson impurity model, embedded into a self-consistently determined hybridization. The imaginary time action of this ``impurity problem''  reads  
\begin{equation}
S^{\mathrm{A}} = -\sum_{i\omega,\sigma} (i\omega + \mu - \Delta_{\omega\sigma}) c^\dagger_{\omega\sigma} c_{\omega\sigma} + S^{\mathrm{int}},
\end{equation}
where $S^{int} = \int_0^\beta d\tau H^{\mathrm{int}} [c^\dagger_i(\tau), c_i(\tau)] $ is the interaction part of the action and $\Delta_{\omega\sigma}$ is a self-consistently determined hybridization function. 
The DMFT impurity solver computes the one particle Green's function $g_{\omega\sigma} = -\langle c_{\omega\sigma} c^\dagger_{\omega\sigma} \rangle$ of the of the Anderson impurity model and the two particle vertex functions (i.e. the connected parts of two-particle Green's functions)  
\begin{equation}\label{eqn:vertex}
\gamma_{\Omega\omega\omega'}^{\sigma_1\sigma_2\sigma_3\sigma_4} = \left(\langle c_{\omega,\sigma_1} c^\dagger_{\Omega + \omega,\sigma_2} c_{\omega' + \Omega, \sigma_3} c^\dagger_{\omega', \sigma_4}\rangle - g_{\omega\sigma_1}g_{\omega'\sigma_3}\delta_{\Omega,0}\delta_{\sigma_1,\sigma_2} + g_{\omega\sigma_1} g_{\omega + \Omega, \sigma_2} \delta_{\omega,\omega'}\delta_{\sigma, \sigma_3} \right).
\end{equation}

The following quantities are then provided as an input to the DF simulation:
\begin{itemize}
\item $g_\omega$ - the full Green's function of the DMFT impurity problem (same values for both spin components)
\item $\Delta_{\omega}$ - hybridization function of the DMFT impurity problem
\item $\mu$ - chemical potential of the problem
\item Two independent components of the impurity vertex function, $\gamma_{\Omega\omega\omega'}^{\sigma_1\sigma_2\sigma_3\sigma_4}$, from Eqn~\eqref{eqn:vertex}: $\gamma_{\Omega,\omega,\omega'}^{\uparrow\uparrow\uparrow\uparrow} \equiv \gamma^{\uparrow\uparrow}_{\Omega,\omega,\omega'} $ and $\gamma_{\Omega,\omega,\omega'}^{\uparrow\downarrow\downarrow\uparrow} \equiv  \gamma^{\uparrow\downarrow}_{\Omega,\omega,\omega'} $.
%\item Lattice dispersion $\varepsilon_{k}$ and the dimensionality $D$ are compiled into the specific running executable. The hypercubic lattice dispersion $-2t \sum_{i=1}^d \cos k_i$ at dimensions $D=1,2,3,4$ is provided with the code. Other lattice choices can be achieved by extending the code. 
\end{itemize}

The present version of the code considers only spin-symmetric solutions of fermionic spin $s=1/2$ problems, and does not describe symmetry-broken phases. We will omit the spin index $\sigma$ in single particle quantities in what follows.

\subsection{Ladder dual fermion self-consistency loop}
A precise derivation of the DF equations can be found in \cite{Hafermann2012, Antipov2014}. Here we outline the equations solved within the \texttt{opendf} code. The evaluation of the DF equations  starts with the construction of the bare dual fermion propagator
\begin{equation}
\tilde G^{(0)}_{\omega,k} = \left[g_{\omega}^{-1} + \Delta_\omega - \varepsilon_k\right]^{-1} - g_{\omega}, \label{eq:gd0}
\end{equation}
which represents a $k$-dependent correction to the impurity Green's function. This Green's function is used to construct two-particle bubbles: 
\begin{equation}\label{eq:dual_bubble}
\tilde \chi_{\Omega\omega}(q) = -\frac{T}{N_k^D} \sum_k \tilde G_{\omega, k} \tilde G_{\omega + \Omega, k+q}.
\end{equation}
Here the integral over the Brilloin zone is replaced with a discrete summation with $N_k$ points in each direction. The impurity vertex functions are combined into density and magnetic channels (labeled d/m respectively) as: 
\begin{equation}\label{eq:spin_symm}
\gamma^{d/m}_{\Omega,\omega,\omega'} = \gamma^{\uparrow\uparrow}_{\Omega,\omega,\omega'} \pm \gamma^{\uparrow\downarrow}_{\Omega,\omega,\omega'}.
\end{equation}
The vertices for the respected channels from Eq. \ref{eq:spin_symm} and the bubbles from Eq. \ref{eq:dual_bubble} are substituted into ladder equations:
\begin{equation}\label{eq:dual_ladder}
\Gamma^{d/m}_{\Omega,\omega,\omega'}(q) = \gamma^{d/m}_{\Omega,\omega,\omega'} + \sum_{\omega''} \gamma^{d/m}_{\Omega,\omega,\omega''} \tilde\chi_{\Omega,\omega''}(q) \Gamma^{d/m}_{\Omega'',\omega'}(q).
\end{equation}
 $\Gamma_{\Omega,\omega,\omega'}$ is called the fully dressed vertex function.

Evaluation of Eq. \ref{eq:dual_ladder} is performed independently for each pair of bosonic frequencies $\Omega$ and transfer momenta $q$. 
$\gamma_{\Omega,\omega,\omega'}$ and $\Gamma_{\Omega,\omega,\omega'}(q)$ are represented as matrices in the space of fermionic Matsubara frequencies $\omega$, $\omega'$, and $\tilde\chi_{\Omega,\omega''}(q)$ is a diagonal matrix. 
In this matrix notation, Eq.~\ref{eq:dual_ladder} reads
\begin{equation}\label{eq:lin_alg_inv}
(\hat 1 - \hat \gamma \tilde \chi)\hat \Gamma  = \hat \gamma.
\end{equation} 
This equation is physically correct only when the maximum eigenvalue of $\hat \gamma \tilde \chi$ is smaller than one, i.e. all eigenvalues of the matrix $\hat D = \hat 1 - \hat \gamma \tilde \chi$ are positive. Eq. \ref{eq:lin_alg_inv} is then solved and $\Gamma$ is obtained. When the determinant of $\hat D$ is negative and a negative eigenvalue exists, the DF solution is outside of the convergence radius of the ladder approximation. Nevertheless, given that the resulting solution is unique, one can extend this convergence radius by doing a low-order iterative evaluation of $\Gamma$ and checking if the inversion of Eq. \ref{eq:dual_ladder} can be obtained on the next DF iteration. 
 
Once the fully dressed vertex function $\Gamma_{\Omega,\omega,\omega'}$ is obtained, it is used in the Schwinger-Dyson equation to obtain the dual self-energy $\tilde \Sigma_{\omega, k}$. 
The equation reads:
\begin{equation}\label{eq:sd}
\tilde \Sigma_{\omega, k} = \frac{T}{2 N_k^D}  \sum_{\Omega, q} \left( 3 \left[\Gamma^m_{\Omega,\omega,\omega}(q) - \frac{1}{2}\Gamma^{(2), m}_{\Omega,\omega,\omega}(q) \right] + \Gamma^d_{\Omega,\omega,\omega}(q) - \frac{1}{2}\Gamma^{(2), d}_{\Omega,\omega,\omega}(q)  \right) \tilde G_{\omega, k + q},
\end{equation}
where $\Gamma^{(2)} = \hat \gamma \tilde \chi \hat \gamma $ indicates the second order (first iteration) correction from Eq. \ref{eq:dual_ladder} to avoid diagrammatic double counting. 

The resulting dual self-energy is used to obtain the dual Green's function from the Dyson equation:
\begin{equation}\label{eq:dyson}
\tilde G^{-1}_{\omega k} = \left[G^{(0}_{\omega k}\right]^{-1} - \tilde \Sigma_{\omega k}
\end{equation}

The procedure is repeated until convergence of $\tilde G$ is achieved. 

\subsection{Resulting observables}
The fully converged dual Green's function $\tilde G$, self-energy $\tilde \Sigma$, vertices $\Gamma^{d/m}$ determine the lattice correlators. Specifically, 
\begin{itemize}
\item the lattice self-energy: 
\begin{equation}\label{eq:sigma_lat}
\Sigma_{\omega, k} = \frac{\tilde \Sigma_{\omega, k}}{1 - g_\omega \Sigma_{\omega, k}} + \Sigma^{DMFT}_{\omega},
\end{equation}
where $\Sigma^{DMFT}_{\omega} = i\omega + \mu - \Delta_{\omega} - g_\omega^{-1}$. 
\item The lattice Green's function
\begin{equation}\label{eq:glat}
G_{\omega,k} = \left[\Delta_{\omega} - \varepsilon_{k}\right]^{-1} + \left[\Delta_{\omega} - \varepsilon_{k}\right]^{-1} g_{\omega}^{-1} \tilde G_{\omega, k} g_{\omega}^{-1} \left[\Delta_{\omega} - \varepsilon_{k}\right]^{-1}.
\end{equation}
Eqs. \ref{eq:sigma_lat} and \ref{eq:glat} are related by a Dyson equation for $G$ and $\Sigma$. 
\item The charge and spin susceptibilities
\begin{align}
\chi^{\mathrm{ch/sp}} (\Omega, q) & = -T \sum_{\omega k} G_{\omega k} G_{\omega+\Omega k+q} + \sum_{\omega,\omega'} L_{\Omega, \omega}(q) \Gamma^{d/m}_{\Omega,\omega,\omega'}(q) L_{\Omega, \omega'}(q), \\
L_{\Omega, \omega}(q) & = -T \sum_k \mathfrak{G}_{\omega,k} \mathfrak{G}_{\omega + \Omega,k + q}, \\
\mathfrak{G}_{\omega,k} & = \tilde G_{\omega k} \frac{\tilde G^{(0)}_{\omega, k} + g_\omega}{\tilde G^{(0)}_{\omega, k}}.
\end{align}
\end{itemize}

\section{Distribution}\label{sec:imp}
The dual fermion code is distributed as a C++ library with compiled executables \texttt{hub\_df\_cubic{\bf D}d}, where \texttt{\bf D} labels the number of dimensions ($D=1,2,3,4$). We use the opensource \texttt{gftools} library \cite{gftools} for algebraic operations with single- and multi-particle Green's functions and its interface to the \texttt{ALPSCore} libraries \cite{ALPSCore} for loading/saving \texttt{hdf5} objects. The code and the documentation are available as Ref.~\cite{OpenDFZenodo}.

\section{Example I and performance analysis}\label{sec:perf}
As a first example and illustration of the performance of the code, we provide an example input generator for the particle-hole symmetric Hubbard model at $U \gg t$ (the  ``atomic limit'') . In this case the input quantities are given analytically by:
\begin{align}\label{eq:atomic_limit1}
g_{\omega} & = \frac{1}{2}\left[\frac{1}{i\omega - U/2} + \frac{1}{i\omega + U/2}\right], \\
\Delta_{\omega} & = 2Dg_{\omega}, \label{eq:atomic_limit2} \\ 
\gamma^{\uparrow\uparrow}_{\Omega,\omega,\omega'} & = \frac{\beta U^2}{4} (\delta_{\omega_1,\omega_2} - \delta_{\omega_1,\omega_4} ) \Lambda_{\omega_1}\Lambda_{\omega_3}, \label{eq:atomic_limit3} \\ 
\label{eq:atomic_limit4}
\gamma^{\uparrow\downarrow}_{\Omega,\omega,\omega'} & = -U + 
\frac{U^3}{8}\frac{\omega_1^2 + \omega_2^2 + \omega_3^2 + \omega_4^2}{\omega_1^2\omega_2^2\omega_3^2\omega_4^2} + \frac{3U^5}{16}\frac{1}{\omega_1\omega_2\omega_3\omega_4}  \\
& \notag + \frac{\beta U^2}{4} \frac{1}{1 + \exp(\beta U /2 )} 
(2\delta_{ \omega_2, -\omega_3} + \delta_{\omega_1, \omega_2}) 
\Lambda_{\omega_2} \Lambda_{\omega_3}  \\ 
& \notag - \frac{\beta U^2}{4} \frac{1}{1 + \exp(-\beta U /2 )} 
(2\delta_{ \omega_1, \omega_4} + \delta_{\omega_1, \omega_2}) 
\Lambda_{\omega_1} \Lambda_{\omega_3},
\end{align}
where $\Lambda_\omega = 1 + U^2/(4\omega^2)$ and  $\omega_1 = \omega, \omega_2 = \omega + \Omega, \omega_3 = \omega' + \Omega, \omega_4 = \omega'$ is used to simplify the notation. The corresponding program is provided with the code. 

The numerical solution of dual fermion equations requires introducing several control parameters. In particular,
the vertex function $\gamma_{\Omega,\omega,\omega'}$ is sampled on a grid with a cutoff $N_\Omega$ in bosonic and $N_{\omega}$ fermionic frequencies and the Brilloin zone is sampled on a finite grid of size $N_k$, giving a total volume of the system of $N_k^D$. We analyze the convergence of the code upon tuning $N_{\Omega}$, $N_{\omega}$ and $N_k$ and the computational effort below. Eqs. \ref{eq:atomic_limit1}, \ref{eq:atomic_limit2}, \ref{eq:atomic_limit3}, \ref{eq:atomic_limit4} are used to provide the input to the code and the system is evaluated in $2$ dimensions, at $U=20$, $\mu = \frac{U}{2}$, $\beta = 1$. We choose the value of $g= \tilde G_{i\pi / \beta, 0, 0}$ to control the convergence. We then plot the normalized difference 
\begin{align}\label{eqn:deltag}
\delta_{g} = \left|\frac{g_{N_x} - g_{N_x \to \infty}}{g_{N_x \to \infty}}\right|
\end{align}
 as a function of control parameter $N_x$, with $x = \{ \Omega, \omega, k \}$, and extrapolate  $N_x \to \infty$ 
to evaluate the error. For the most expensive point shown here, the run-time of the simulation was $\approx 2$ min on a laptop.

\begin{figure}[ht]
\includegraphics[width=1.0\columnwidth]{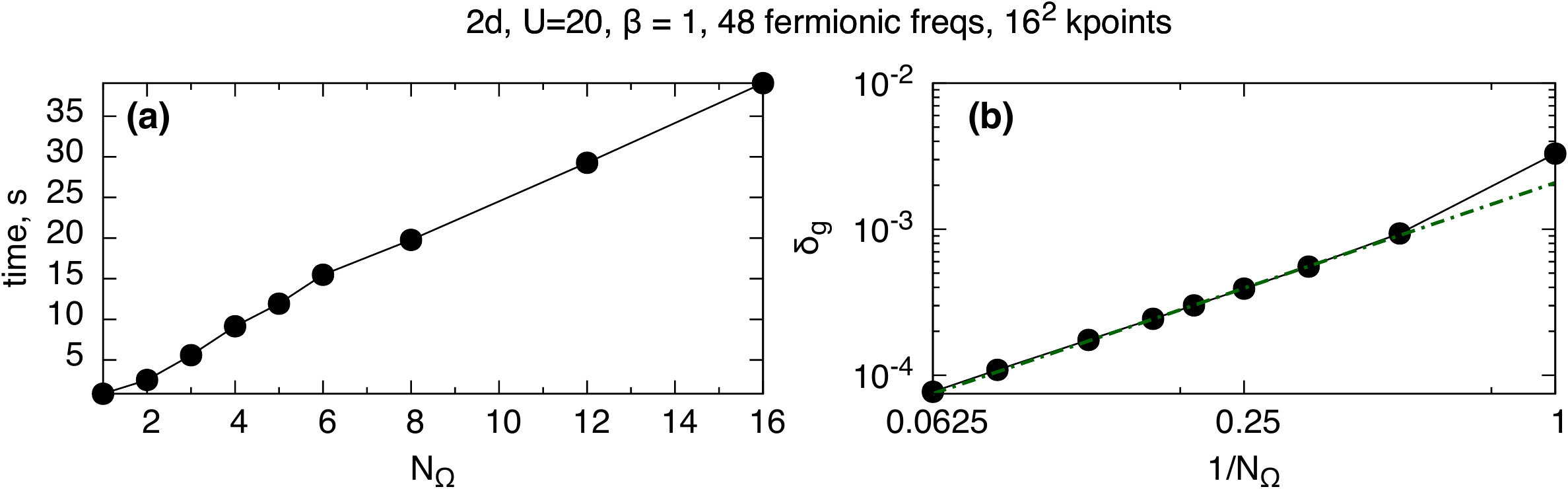}
\caption{(a) Execution time of the dual fermion calculation for the Hubbard model in $2$ dimensions with ``atomic limit'' input at $U=20$, $\beta = 1$ as a function of the number of bosonic frequencies $N_{\Omega}$ at $N_{\omega} = 48,~N_k = 16$; (b) Systematic error $\delta_g$ of the dual fermion Green's function $\tilde G_{i\omega, k}$ at $i\omega = i\pi / \beta, k = (0,0)$ as a function of bosonic frequencies $N_\Omega$, plotted on a logarithmic scale. }
\label{fig:benchmark_b}
\end{figure}

Fig. \ref{fig:benchmark_b} shows the performance of the \texttt{opendf} code upon the change of the total number of bosonic frequencies $N_{\Omega}$ in the vertex $\gamma_{\Omega}$ for a fixed number of fermionic frequencies $N_{\omega}=48$ for a $16 \times 16$ k-space grid. The computational effort, indicated by the time to convergence in Fig.~\ref{fig:benchmark_b}(a), grows linearly in $N_{\Omega}$. The error $\delta_g$, as defined in Eqn.~\ref{eqn:deltag} and shown in frame (b), is of the order of a percent and decreases with a power law.

\begin{figure}[ht]
\includegraphics[width=1.0\columnwidth]{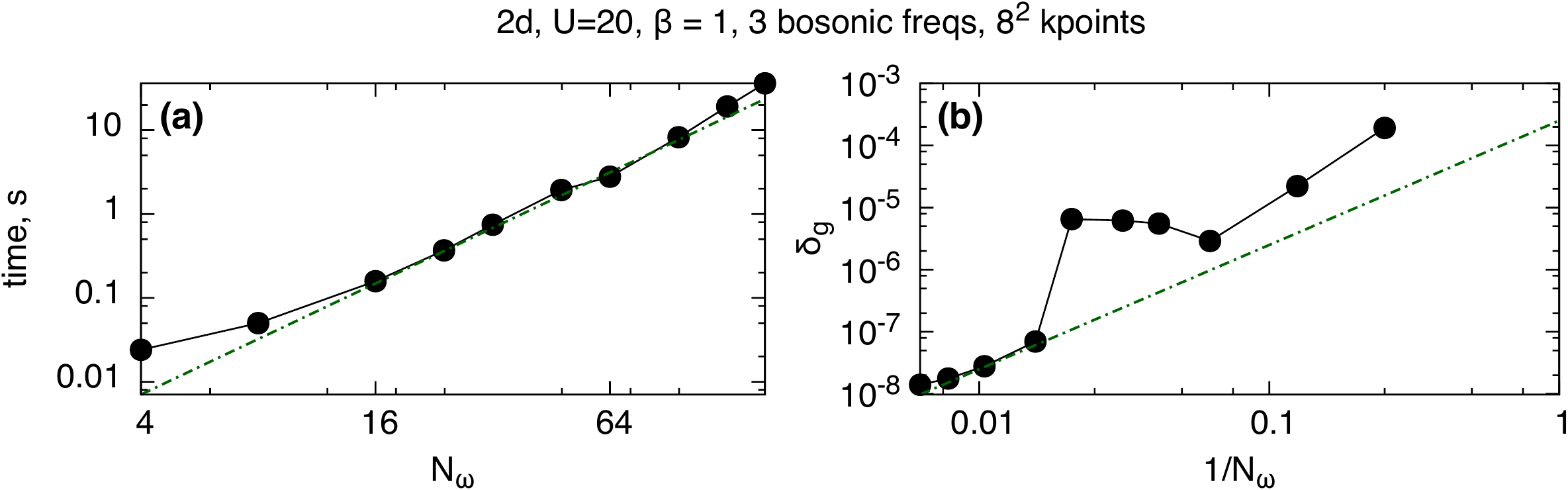}
\caption{(a) Execution time of the dual fermion calculation for the Hubbard model in $2$ dimensions with ``atomic limit'' input at $U=20$, $\beta = 1$ as a function of the number of fermionic frequencies $N_{\omega}$ at $N_{\Omega} = 3$, $N_k = 8$; (b) Error $\delta_g$ of $G_{i\omega, k}$ at $i\omega = i\pi / \beta, k = (0,0)$ as a function of $1/N_{\omega}$, for the same parameters.}
\label{fig:benchmark_f}
\end{figure}

We analyze the performance of the code with respect to the change of the total number of fermionic frequencies $N_{\omega}$  in Fig. \ref{fig:benchmark_f}.  In this benchmark we fix the number of bosonic frequencies, $N_{\Omega}=3$, and perform the calculation on a $8\times8$ k-space grid. The computational expense seen in Fig.~\ref{fig:benchmark_f}(a) grows almost quadratically, while the relative error shown in Fig.~\ref{fig:benchmark_f}(b) is an order of magnitude smaller, as compared to the variation in $N_{\Omega}$ shown in Fig.\ref{fig:benchmark_b}(b) and reduces as a power-law with an increase of $N_{\omega}$.

\begin{figure}[ht]
\includegraphics[width=1.0\columnwidth]{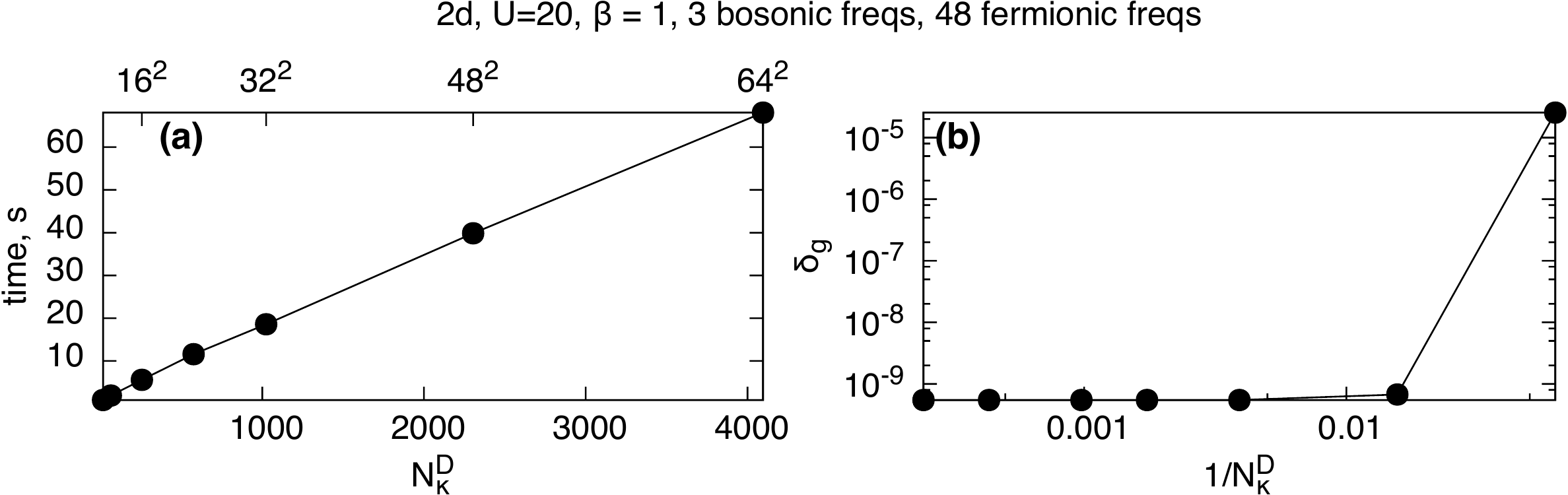}
\caption{(a) Execution time of the dual fermion calculation for the Hubbard model in $2$ dimensions with ``atomic limit'' input at $U=20$, $\beta = 1$ as a function volume $N_k^D$ at $N_{\Omega} = 3$, $N_{\omega} = 48$. (b) Systematic error $\delta_g$ of $G_{i\omega, k}$ at $i\omega = i\pi / \beta, k = (0,0)$ as a function of the volume at the same parameters.}
\label{fig:benchmark_kpts}
\end{figure}

The performance of the code with respect to the change of number of k-space samples within the Brilloin zone $N_k^D$, is plotted on Fig. \ref{fig:benchmark_kpts}. The computational effort (frame (a)) scales linearly with the volume $N_k^D$ of the system and shows fast convergence of the relative error $\delta_g$ (frame (b)).

\section{Example II - Hubbard model, 2 dimensions}\label{sec:ex}

\begin{figure}[ht]
\centering
\includegraphics[width=0.6\columnwidth]{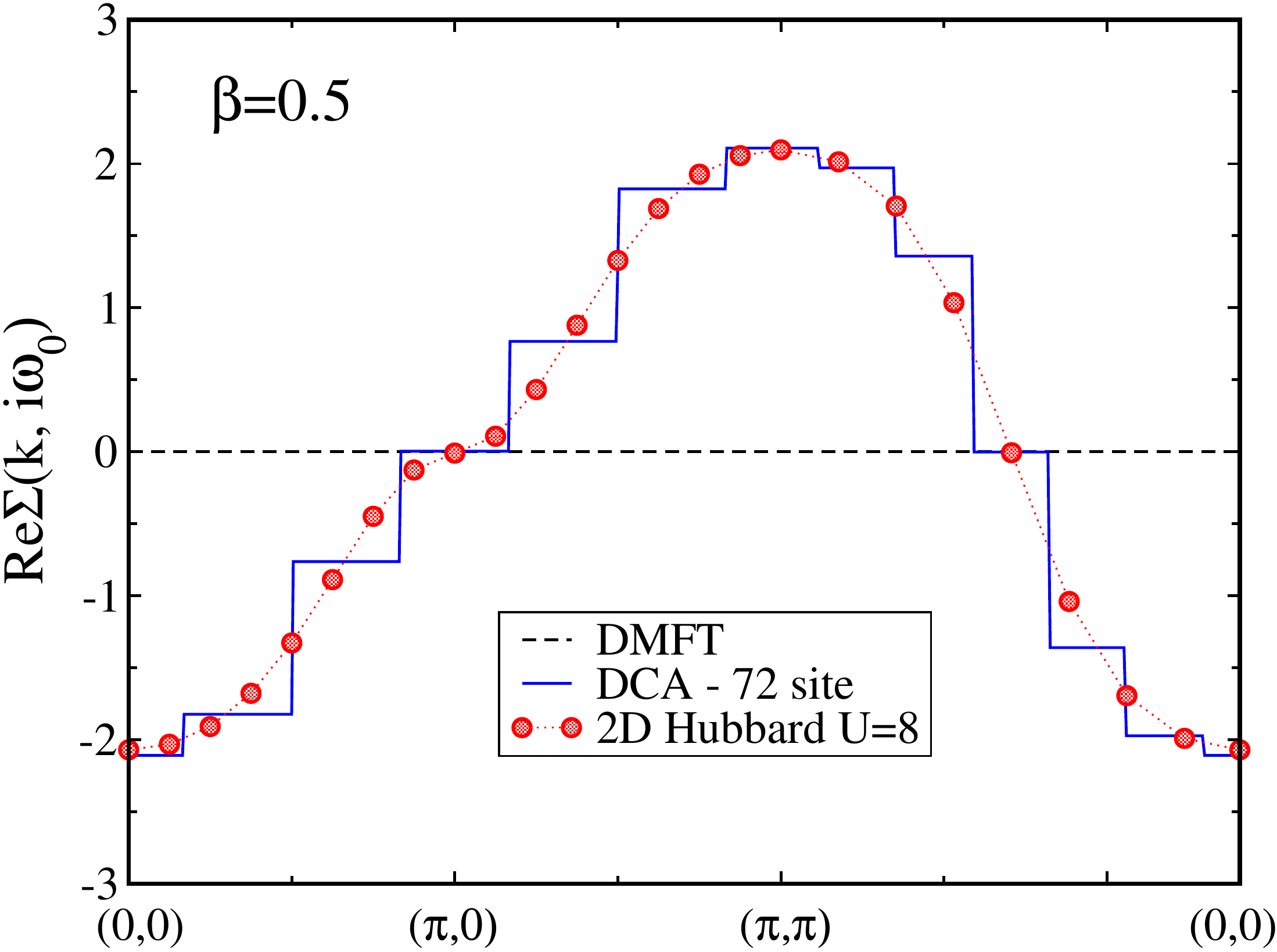}
\caption{Momentum dependence of the real part of the lower Matsubara frequency of the lattice self-energy of the particle-hole symmetric Hubbard model in $2$ dimensions, as obtained by the \texttt{opendf} calculation at $U/t = 8$, $\beta = 0.5$ along the $(k_x,k_y) = (0,0)\to(\pi,0)\to(\pi,\pi)\to(0,0)$ path (red points). Shown also is the DMFT value (as a dashed line) and the comparison data from the $72$-site Dynamical Cluster Approximation calculation (solid blue line).}
\label{fig:sigma}
\end{figure}

\begin{figure}[ht]
\centering
\includegraphics[width=\columnwidth]{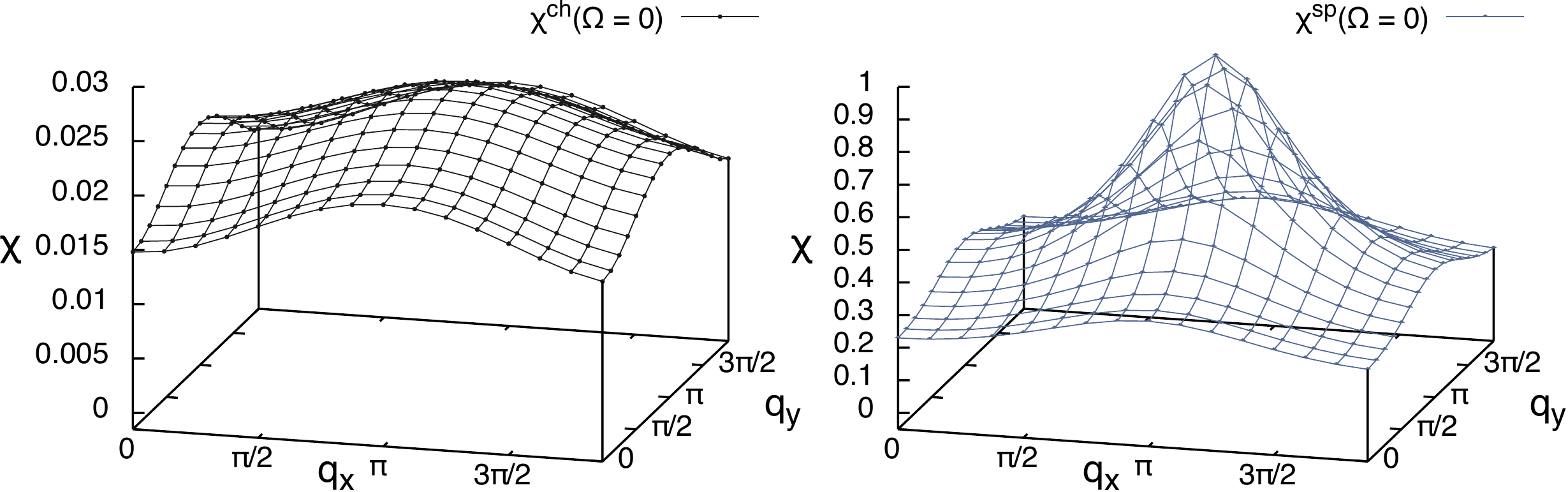}
\caption{$\chi^{\mathrm{ch}}$ (left panel) and $\chi^{\mathrm{sp}}$ (right), the static spin and charge susceptibilities, at $\Omega = 0$ as a function of momenta $q_x$ and $q_y$ as obtained by the \texttt{opendf} calculation for $\beta = 0.5$, $U/t = 8$. }
\label{fig:susc}
\end{figure}

We provide the practical illustration of the method for the Hubbard model in $D=2$ dimensions. We show the $k$-dependence of the real part of the lattice self-energy $\Sigma(k, i\omega_n)$ at $i\omega_n=i\omega_0 \equiv i\pi/\beta$ in Fig.~\ref{fig:sigma} for the case of particle-hole symmetry at $U/t=8$ and compare it with available data from the Dynamical Cluster Approximation \cite{MaierJarrell:2005}. The impurity model, solved using the ALPS DMFT \cite{ALPSDMFT} package with a CT-AUX solver \cite{Gull2008b}, was used as an input. The DMFT self-energy is momentum-independent, $\mathrm{Re} \Sigma_{\omega k}^{\mathrm{DMFT}} = 0$, and is plotted with a dashed line. Taking into account the spatially dependent corrections by the dual fermions leads to a correct momentum-dependence of the self-energy, matching in this case the DCA result. A detailed comparison between multiple methods will be discussed elsewhere \cite{LeBlanc2015}. 

We illustrate the susceptibility in Fig. \ref{fig:susc}. Plotted are the static spin- and charge- susceptibilities at $U/t = 8$ for the particle-hole symmetric case. The spin susceptibility, peaked at $(\pi,\pi$) due to antiferromagnetic fluctuations is much larger than the charge one. 

\section{Conclusion}\label{sec:conclusions}
In this paper we have introduced an open source implementation of the dual fermion method, the \texttt{opendf} project. It solves the dual fermion self consistency equations and computes non-local corrections to the local solutions provided by DMFT.  \texttt{opendf} can be used to augment DMFT computations with two-particle quantities and add momentum dependence to DMFT observables. 

Future development of the code is anticipated. Further releases will include extensions to additional diagrams, broken-symmetry phases and multi-orbital systems. 

\section*{Acknowledgements}
We are grateful to D. Hirschmeier for fruitful discussions and acknowledge the Simons collaboration on the many-electron problem for financial support and for its support of the \texttt{ALPSCore} project.

\bibliography{opendf}

\end{document}